# Predicting Locations of Cell Towers for Network Capacity Expansion

Sowmiyan Morri, Joy Bose, L Raghunatha Reddy, Sai Hareesh Anamandra
Ericsson
Bangalore, India
joy.bose@ericsson.com

*Abstract*— Network capacity expansion is a critical challenge for telecom operators, requiring strategic placement of new cell sites to ensure optimal coverage and performance. Traditional approaches, such as manual drive tests and static optimization, often fail to consider key real-world factors including user density, terrain features, and financial constraints. In this paper, we propose a machine learning-based framework that combines deep neural networks for signal coverage prediction with spatial clustering to recommend new tower locations in underserved areas. The system integrates geospatial, demographic, and infrastructural data, and incorporates budget-aware constraints to prioritize deployments. Operating within an iterative planning loop, the framework refines coverage estimates after each proposed installation, enabling adaptive and cost-effective expansion. While full-scale simulation was limited by data availability, the architecture is modular, robust to missing inputs, and generalizable across diverse deployment scenarios. This approach advances radio network planning by offering a scalable, data-driven alternative to manual methods.

## I. Introduction

The rapid expansion of mobile communication networks has created a pressing need for scalable and intelligent methods of radio network planning. One of the most critical challenges faced by telecom operators is determining the optimal placement of new cell sites to ensure adequate coverage and meet increasing demand. Traditional methods for network expansion such as drive tests, manual surveys, and rule-based heuristics are not only time-consuming and resource-intensive, but often fail to account for important real-world constraints, including geographical features, user demand variability, and financial limitations.

Existing literature has explored a range of optimization-based and heuristic techniques for base station placement, yet many rely on simplified assumptions that do not generalize well in practical deployments. Assumptions such as uniform user distribution or fixed terrain conditions ignore the inherent complexity of urban and rural environments. Moreover, ad hoc approaches currently used in the industry—such as signal latching, intra-circle roaming, and frequency reuse—may yield marginal improvements but often fall short in sustaining long-term network performance.

In practice, the process of site selection is further complicated by a combination of technical and business factors. Decisions about where and how many towers to install are typically made through consultations between network engineering teams and commercial departments, factoring in customer acquisition forecasts, budget allocations, and operational feasibility. Misalignment between technical recommendations and business expectations can lead to under-utilized infrastructure or unmet service quality targets.

This paper addresses these challenges by proposing an AI-driven framework for data-informed and budget-aware network expansion. At the core of our methodology is a deep learning model trained to predict signal coverage based on multiple input features, such as distance to existing towers, terrain, urban density, and antenna characteristics. The predicted coverage maps are used to identify low-coverage regions, which are then clustered spatially to generate candidate locations for new tower installations. A budget-aware decision module filters and ranks these candidates, ensuring that deployment plans are both technically effective and financially viable.

Unlike previous approaches, our system integrates predictive modelling, spatial analytics, and cost-constrained optimization into a unified, iterative loop. This allows telecom operators to make adaptive decisions informed by data, thereby aligning network planning more closely with real-world dynamics. Though our simulation was limited by data and infrastructure constraints, the framework's modular structure and generalizability make it suitable for deployment across diverse geographical and economic contexts.

Through this work, we aim to bridge the gap between theoretical planning models and operational realities, offering a scalable, adaptive, and explainable tool for next-generation telecom network expansion.

The remaining sections of the paper are as follows: section 2 explains the problem statement. Section 3 contains related work in the area of coverage expansion. Section 4 details our proposed AI driven methodology, including details of the data collection, ML modules, and iterative optimization. Section 5 explains our proposed system architecture in detail. Section 6 expands on the budgetary constraints in real-world deployments. Section 7 summarizes our difficulties for simulation. Section 7 includes a short discussion, and section 9 concludes the paper.

## II. Problem Statement

If a team of a telecom operator installs new sites without considering the recommendations of the sales department, they may be held responsible for poor sales performance. Therefore, it is crucial to have an accurate estimate of both the customer utilization and the number of potential customers in a new area before deciding to install new sites.

Currently, telecom operators use several ad hoc techniques to enhance network coverage and capacity, including latching, intra-circle roaming (for example, from Airtel to VI), and half cycle methods to accommodate more calls. The general procedure for network enhancements involves the following:

- Performing a drive test to collect data on network performance.
- Analyzing the results of the drive test by a team of engineers.
- Making recommendations based on the analysis.
- Consulting with the sales team to align technical decisions with business goals.

When planning to install new sites in a previously unserved area, operators typically follow the following steps:

- Obtain geographical map data of the target area, although this step can be expensive.
- Conduct virtual studies comprising multiple iterations, optimizing site locations and cell frequencies.
- Develop and recommend a comprehensive site installation plan.

From the viewpoint of the Network Engineer (NE), even after all optimizations and tests, poor coverage can persist in certain areas. Furthermore, decisions to install new towers are influenced by various external factors such as budget constraints and the sales team's interest. Typically, the top management sets the budget and limits the number of sites (say N) to be installed in a particular area. These new sites are generally installed for coverage enhancement or capacity enhancement, proposed by the relevant operational teams after incorporating feedback from the sales department.

This approach aligns technical network planning with expected business outcomes because if new sites do not generate sufficient utilization or increase the customer base in the initial months, the investment is considered unjustified. Conversely, if sites are installed without proper alignment to business expectations, it may lead to underperformance in sales metrics.

We propose a method for radio network planning that is supported by simulation results. It evaluates the field strength at every point within a specified area. However, the simulations rely on certain assumptions and constraints that may not be applicable in real-world scenarios. For instance, one key assumption is that each base station serves an equal number of users. Since the study focuses on optimizing base station locations and transmit power under this assumption, the results may not be practical for actual implementations. Consequently, the analysis is primarily theoretical. Additionally, the optimization outcomes presented may represent local optima rather than the global optimum. The study also does not take into account important factors such as geographical features like water bodies, buildings, and forest cover, which can significantly affect network coverage.

From the perspective of telecom operators, installing base stations without considering inputs from the sales department could lead to poor sales performance. Therefore, it is important to accurately estimate user demand and population distribution in new areas before planning network expansion. Current ad hoc techniques used to improve coverage and capacity include methods like latching, intra-circle roaming, and half-cycle techniques. The typical procedure for network enhancement involves conducting drive tests, analyzing the resulting data by a team of engineers, making recommendations, and consulting with the sales team.

## III. Related Work

The problem of mobile network capacity planning, particularly the optimal placement of base stations (BS) and cell towers has received considerable attention in both academic literature and industry practice. Various approaches have been proposed over the years, ranging from heuristic-based techniques to analytical optimization models and, more recently, machine learning-driven methods. However, most existing solutions suffer from practical limitations when applied to real-world deployments, particularly in heterogeneous environments where user demand, geographic features, and financial constraints vary significantly.

One stream of research has focused on the use of reinforcement learning for dynamic and adaptive infrastructure deployment. For example, Qiu et al. [1] proposed a deep reinforcement learning-based approach for optimizing the placement of aerial base stations (ABS) in unmanned aerial vehicle (UAV) networks. Their model dynamically adjusts ABS positions to maximize coverage, which is particularly effective in temporary or mobile communication setups. However, their approach is less suited to terrestrial cell tower infrastructure, where installations are fixed and long-term in nature.

A more traditional line of work was advanced by Binzer and Landstorfer [2], who introduced a neural network-based model for radio network planning. Their approach simulated coverage across a target region based on simplified assumptions, including uniform user distribution and equal load across base stations. While the method provided useful insights into radio propagation modelling, it neglected critical real-world factors such as budgetary constraints, terrain features (e.g., buildings, water bodies), and potential subscriber growth in rapidly developing areas. Furthermore, the solutions produced by their optimization model were typically local optima, limiting their effectiveness in dynamic or large-scale deployments.

From an industrial perspective, a patent by China Mobile (US10477413B2) [3] proposed a rule-based system for capacity expansion by monitoring online traffic data across multiple network cells. Their method relies on identifying service types, traffic loads, and usage durations to assess infrastructure demands. However, this solution neither incorporates predictive modelling nor determines optimal tower locations. Additionally, it is based on fixed thresholds and lacks adaptability to shifting demand or environmental changes.

A paper by Bharadwaj et al. [4] leverages high-resolution 3D terrain data for highly precise signal strength prediction and subsequent optimal tower placement. Another work [5] employs an improved mean shift clustering algorithm within a multi-step framework for 5G base station deployment, optimizing for cost and addressing weak coverage areas.

There is some recent research that uses variants of reinforcement learning to optimize coverage and select base locations. A work [6], involving an AI framework for 5G base station site selection integrates Large Language Models (LLMs) for user interaction and Reinforcement Learning (RL) using Group Relative Policy Optimization (GRPO) for multi-objective optimization. The paper by Al-Tahmeesschi et al. [7] uses Deep Reinforcement Learning (DRL) with a Deep Q-Network (DQN) for the joint optimization of coverage rate and localization accuracy in 5G urban environments.

What distinguishes our proposed framework from these earlier efforts is its integration of data-driven coverage modelling, spatial clustering, and cost-aware optimization into a unified planning loop. Unlike approaches that assume uniformity or ignore economic feasibility, our method leverages a deep neural network to capture complex relationships among infrastructure parameters, geographic features, and observed coverage quality. Moreover, by incorporating clustering algorithms to identify underserved regions and using budget constraints to prioritize deployment, the proposed system aligns technical planning with real-world business considerations.

In summary, while prior studies have contributed valuable insights, ranging from theoretical models to operational heuristics, they often overlook essential constraints or fail to generalize across diverse deployment scenarios. Our work seeks to bridge this gap by developing a robust, flexible, and scalable framework that supports informed decision-making for network expansion across both urban and rural contexts.

## IV. PROPOSED METHODOLOGY

In this section, we describe the detailed methodology employed for optimizing the placement of new cell towers to enhance network coverage and capacity in a given area. Our approach integrates iterative coverage evaluation, clustering techniques, and KPI-driven optimization to recommend optimal site locations for network expansion.

The objective of our proposed methodology is to develop a machine learning-driven framework for automated radio network planning, particularly for identifying optimal locations for new cell sites (or towers) in order to improve coverage and capacity in a designated area. Our approach addresses the limitations of manual and semi-automated planning techniques by incorporating spatial analytics, clustering algorithms, and predictive modelling, all while accounting for real-world constraints such as geography, user demand, and budget.

The implementation of our proposed solution is based on an AI/ML-driven system designed to automate and optimize the network expansion process for telecom operators. It integrates deep learning-based coverage prediction, clustering algorithms for candidate site selection, and budget-aware iterative optimization.

**Problem Setup and Assumptions**

We model the target geographical area as a discrete 2D grid of size n*n, resulting in $n^2$ candidate points. Each point may or may not currently contain a cell site. This setup is encoded as a binary vector $x \in \{0,1\}^{n^2}$, where $x_i=1$ indicates the presence of a cell site at location i, and 0 indicates absence.

Associated with each point is a real-valued coverage indicator $c_i$, such as RSSI or CQI, forming a vector $c \in R^{n^2}$. The objective is to identify new locations where placing additional towers maximally improves overall coverage, within a given deployment budget.

We assume access to historical coverage and deployment data, including drive tests, crowdsourced signal measurements, terrain data, population density, and current site configurations. Missing data is expected and handled during preprocessing.

**Data Acquisition and Preparation**

The system utilizes a variety of data sources including:

- Drive test logs and crowdsourced signal measurements (e.g., P3, UMLAUT) for coverage mapping.
- Performance management (PM) counters for network KPIs.
- Geospatial data from public APIs (e.g., Google Maps) for buildings, terrain, and water bodies.
- Demographic data, such as population density and subscriber statistics (both existing and potential).
- Site configurations, including locations of current cell towers and antenna characteristics.

Preprocessing includes handling missing data through:

- Record exclusion for missing independent variables (e.g., terrain).
- Aggregated imputation for dependent variables (e.g., KPI values like RSSI, SINR) using cell-site granularity.

Our framework includes two machine learning modules which are detailed below:

**Machine Learning Module 1: Coverage Prediction Module**

We define a function $f_\theta: R_m \rightarrow R$ parameterized by neural network weights $\theta$, where each input vector consists of m features relevant to coverage prediction:

- Distance to nearest cell towers
- Terrain and altitude
- Urban/rural classification
- Population/user density
- Frequency band, antenna type, and directionality

A deep neural network is trained on the collected features to learn the coverage equation. The trained model generalizes coverage quality across the spatial grid, providing a complete coverage vector over the target area.

The neural network is trained to minimize the error between predicted and observed KPI values (e.g., RSSI) using supervised learning on a labelled dataset $\{(a^{(i)}, c^{(i)})\}$.

This model learns the coverage equation mapping from each location's input vector (e.g., distance to nearest cell, terrain type, frequency band, antenna direction, etc.) to a predicted signal strength (RSSI or SINR):

$c_i = f_\theta(a_{i1}, a_{i2}, \ldots a_{im})$

Once trained, $f_\theta$ can estimate coverage at any new location given its features and current network configuration.

This module supports both inference on new areas and generation of synthetic coverage data for simulations. The DNN architecture was chosen over shallow models for its ability to capture non-linear interactions between features, especially in complex terrain.

**Machine Learning Module 2: Cell Location Recommender**

This module identifies low-coverage regions where predicted signal strength falls below a threshold. Spatial clustering algorithms (DBSCAN or K-means) are then applied to these regions to identify groups of under-served areas. Cluster centroids are proposed as candidate locations for new towers.

Predicted coverage across the grid is analysed to identify low-coverage regions. All points where predicted coverage $c_i$ falls below a predefined threshold $\tau$ are extracted. Clustering algorithms such as DBSCAN or K-means are applied to this subset to group spatially proximate low-coverage points into clusters $C_1, C_2, \ldots C_k$.

The centroids of these clusters are chosen as initial candidate locations for new towers. For each cluster, the potential gain in coverage is evaluated, and clusters are ranked by effectiveness.

The number of candidate towers to be selected is constrained by the total deployment budget B, given a fixed cost per tower y:

Max new sites = $\lfloor B/y \rfloor$

We compare different placement strategies:

- Place one tower at cluster centroid
- Place multiple towers at high-impact points along the cluster's boundary

A greedy selection process is used to choose placements that offer the maximum marginal improvement in predicted coverage.

This process is influenced by budget constraints, which limit the number of towers deployed (see Budget Constraints section). The system iteratively evaluates whether placing towers at centroids or distributing them across cluster peripheries yields better overall coverage improvements.

**Budget Constrained Iterative Optimization**

The network planning process is executed iteratively as an adaptive loop as follows:

1. Predict coverage using current cell site configuration
2. Identify low-coverage points ($c_i < \tau$)
3. Cluster those low coverage points
4. Recommend candidate sites.
5. Add one or more selected sites to the configuration
6. Re-compute the predicted coverage with updated configuration
7. Repeat until:
   o Target coverage metrics are achieved, or
   o Budget B is exhausted

Budget plays a critical role in determining the feasibility of the expansion plan. Our model incorporates cost-awareness by calculating the maximum number of new towers that can be installed within a given budget, using a simple constraint: installed within a given budget, using a simple constraint:

Max Sites = Total Budget/Cost per Site

During clustering, this constraint guides whether fewer towers (e.g., at cluster centroids) or more towers (e.g., at multiple endpoints) would yield a better cost-benefit trade-off. The solution dynamically adjusts the number and placement of towers to meet both technical goals and financial constraints.

This feedback loop ensures that each deployment decision is made in light of the current network state, progressively refining site selection. It ensures that each newly added tower is selected based on the current state of the network and not static heuristics.

This iterative approach allows for adaptive planning, wherein each new site is chosen based on current network conditions and predicted benefits rather than static assumptions.

**Handling Incomplete Data**

In practical settings, data for certain features or KPIs may be missing. We adopt the following strategies:

- If an independent variable (e.g., terrain feature) is missing, we drop the sample
- If dependent variables (e.g., RSSI) are missing, we use interpolation or aggregate neighbourhood-level imputation

This ensures that the DNN can be trained on a clean dataset while preserving the general patterns in the data.

**Deployment Scenarios**

Our model is designed to support a wide range of practical deployment scenarios:

- Urban densification: Suggest additional towers in high-demand but saturated areas
- Rural expansion: Plan initial deployments in underdeveloped, remote or rural areas with sparse population

- Partial upgrade: Augment existing networks in towns
- Highway or corridor expansion: Enhancing connectivity along transit routes.
- Greenfield planning: Recommend first tower locations in completely uncovered zones

Because the model is data-driven and spatially aware, it generalizes well across diverse topographies and population patterns.

**Robustness and Generalizability**

Our system is designed to generalize across varying geographic and demographic settings. Once trained on sufficiently diverse data, the framework can be ported to new regions with similar characteristics. Furthermore, it maintains robustness to missing or incomplete data, a common issue in telecom planning, through imputation and tolerance for partial information. This integrated implementation advances telecom planning from rule-based heuristics toward a systematic, explainable, and cost-effective AI-supported process.

One of the strengths of our approach is its ability to operate under partial data availability. In cases where certain features are missing (e.g., incomplete KPI data or coverage logs), the model is designed to handle such gaps through imputation strategies or generalized learning from similar areas.

Furthermore, the framework is designed to be generalizable across different geographic regions. Once trained on a sufficiently diverse dataset, the model can be applied to new areas with similar terrain and population characteristics, making it useful for both urban densification and rural network expansion scenarios.

**Key Advantages of our approach**

The advantages of our approach are as follows:

- **Data-Driven**: Moves away from manual heuristics to a model-based approach using actual coverage and subscriber data.
- **Scalable**: Can scale across large geographical areas and adjust to real-time demand and constraints.
- **Adaptive**: Iteratively improves planning based on updated coverage metrics after each simulated tower addition.
- **Cost-Aware**: Integrates budget constraints into the site selection process.
- **Geographically Informed**: Accounts for water bodies, buildings, and terrain in coverage estimation.

## V. SYSTEM ARCHITECTURE

To operationalize our proposed methodology for intelligent cell tower placement, we designed a modular AI/ML system architecture that integrates spatial modelling, deep learning, and iterative optimization. The system comprises several core components, each responsible for a distinct stage of the planning process. In this section, we describe the architecture and function of these components, along with illustrative diagrams that support conceptual understanding.

**Grid-Based Spatial Representation**

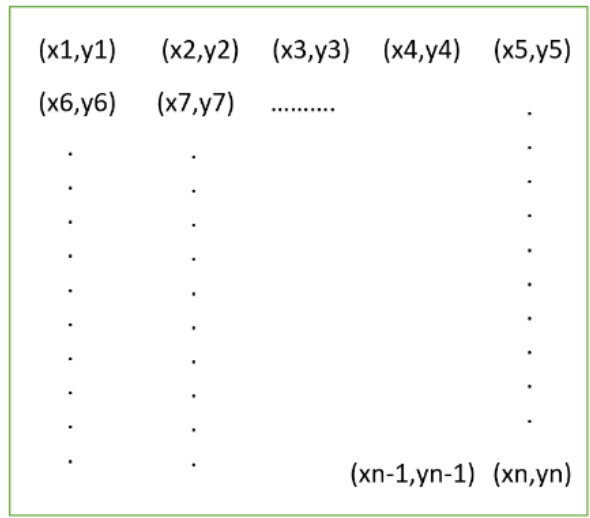

Fig. 1. Visualization of a unit area, broken into n equally spaced points on X and Y dimensions. This is represented by an n-dimensional binary vector [1 0 0 …. 1…0], with 1 standing for a cell currently existing at that point, 0 for a cell (or tower) not existing at that point. This vector we call the cell vector for that area. Similarly, there is a real valued $n^2$ dimensional coverage vector, which gives the coverage at each point in that area.

Fig. 1 shows a visualization of Unit Area and Binary Vectors.

This diagram depicts a unit geographical area. The geographical area is represented as a two-dimensional grid, discretized into n*n cells. Each grid point corresponds to a potential location for a cell tower. The existing network configuration is encoded as a binary vector of length $n^2$, where a value of 1 denotes the presence of a tower at that location and 0 denotes absence. In parallel, a real-valued vector represents coverage strength (e.g., RSSI, SINR) at each point. This dual representation forms the foundation for spatial modeling and enables the downstream coverage prediction and optimization modules.

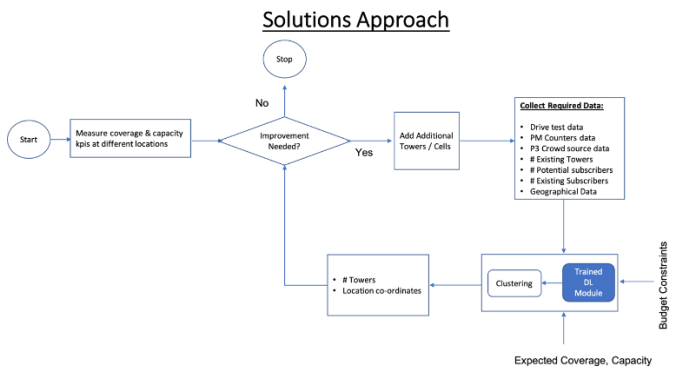

Fig. 2. High level flowchart of the AI/ML solution to determine the optimal locations of the cells.

Fig. 2 shows the high-level system workflow of our AI/ML solution to determine the optimal cell locations.

The system follows a structured, end-to-end workflow that begins with data collection and preprocessing. This includes integrating multiple data sources such as drive test logs, network performance counters, geospatial attributes, demographic statistics, and existing infrastructure configurations. The preprocessed data is then input into the Coverage Prediction Module (Module 1), which employs a trained deep neural network to estimate signal quality at each point in the spatial grid. Regions identified as having poor coverage are subsequently processed by the Cell Location Recommender (Module 2), which applies spatial clustering algorithms to determine optimal locations for new tower installations. This entire pipeline operates in an iterative loop, where each round of predictions and site recommendations leads to an updated network configuration. The process continues until predefined coverage targets are met or the available budget is exhausted, enabling adaptive and cost-aware network expansion.

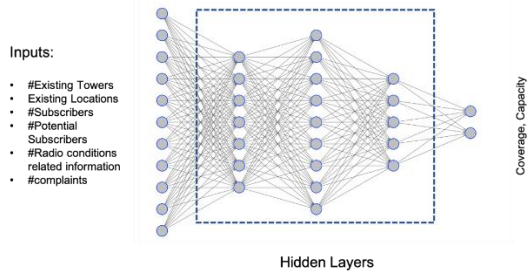

Fig. 3. Architecture of an embodiment of the AI/ML solution to determine the coverage at each of the coordinates in the considered geographical area. Here the AI/ML model is represented as a deep neural network (DNN)

The figure 3 presents the architecture of the deep learning network (DNN) used for coverage estimation. The DNN is trained to predict signal strength at any location within the grid. The input to the DNN includes multiple features for each location including the following:

- Distance to nearest existing towers
- Geographical features (e.g., proximity to water bodies or buildings)
- User or Subscriber density
- Environmental data (terrain elevation, urbanity index, etc.)
- Urban versus rural classification
- Frequency band, antenna type, and orientation

The hidden layers are structured to capture complex nonlinear interactions between these features, leading to an output node representing predicted coverage (e.g., RSSI) at a particular coordinate.

The DNN thus captures non-linear relationships among these features and outputs a predicted signal strength for each location. The model is trained using supervised learning on labelled datasets containing measured KPIs and associated features. Once trained, it enables inference over new or hypothetical configurations, making it suitable for both brownfield and greenfield planning.

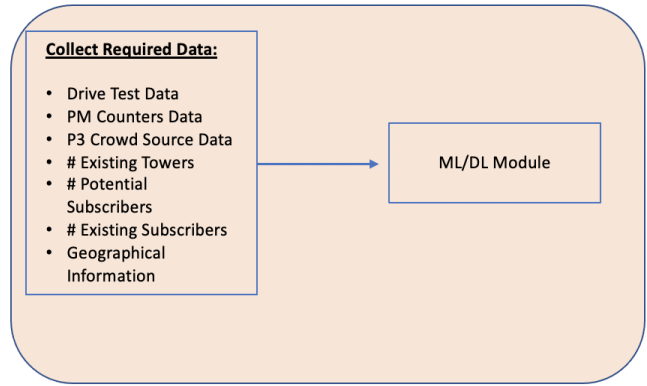

Fig. 4. Illustration of the training of the machine learning model to predict locations of cell towers for network capacity expansion

Fig. 4 illustrates the training workflow for Coverage Prediction. The system uses a labelled dataset comprising locations (lat-long), features (e.g., frequency bands, antenna type), and their corresponding measured KPIs (RSSI, CQI, etc.) obtained from drive tests and crowdsourced platforms like UMLAUT or P3. The model minimizes prediction error between observed and predicted KPI values during training.

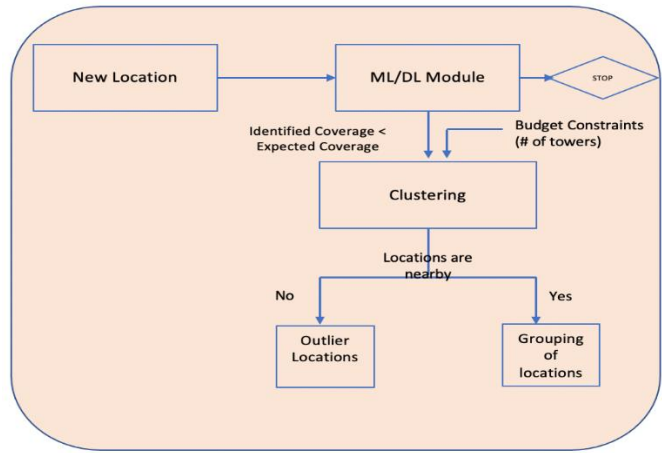

Fig. 5. Illustration of the application phase for inference of the machine learning model to predict locations of cell towers for network capacity expansion

Fig. 5 shows the Inference Workflow for New Site Recommendation. Once the model is trained, it is deployed to predict coverage for a new area using existing site configurations. The predicted coverage vector is analysed to identify under-served areas. Clustering is performed on low-coverage points, and centroids of these clusters are output as suggested new tower coordinates. These candidates are evaluated iteratively to simulate cumulative improvement in coverage.

### Clustering-Based Site Selection Logic

Locations predicted to have poor coverage (below a predefined threshold) are extracted and clustered using spatial algorithms such as DBSCAN or K-means. Each resulting cluster represents a contiguous region of suboptimal signal quality.

The centroid of each cluster is initially proposed as a new site for tower installation. These recommendations are further refined based on potential coverage gain and cost-effectiveness, ensuring that each deployed tower contributes maximally to the network's performance. If the number of candidate clusters exceeds the budgetary limit, a pruning algorithm selects the most impactful clusters based on coverage gain per cost ratio.

### Iterative Refinement Loop

To support progressive deployment, the system operates in an iterative feedback loop. At each step the following computations are performed:

1. The DNN estimates coverage using the current configuration.
2. Low-coverage areas are clustered, and candidate tower sites are generated.
3. A subset of these sites is selected based on cost-benefit trade-offs and added to the configuration.
4. The updated configuration is used for the next round of prediction.

This loop continues until desired performance metrics are achieved or financial constraints are reached. The adaptive nature of this loop allows the system to refine decisions based on real-time conditions, mirroring practical network deployment practices.

Details of the budget-constrained planning are shown in the section 6.

### Integration

The described architecture presents a scalable, modular, and explainable approach to automating cell tower placement. Each component, from grid abstraction to DNN training, clustering, and budget-based filtering, contributes to a holistic solution that is adaptable to diverse geographical and financial contexts. The architecture facilitates an explainable and systematic transition from raw data to actionable deployment strategies. It combines AI-based coverage modelling, spatial analytics, clustering, and budgetary reasoning in a coherent pipeline.

This architecture has the potential to significantly reduce manual effort in telecom network expansion and bring higher precision and responsiveness to the planning process. The modular design allows for future enhancements such as incorporating time-series modelling or reinforcement learning without disrupting the core functionality. As such, it offers a robust foundation for next-generation telecom planning.

## VI. Budget Constrainted Planning in Deployments

In real-world telecom deployments, financial limitations are a critical constraint that influence both the number and placement of new cell towers. Unlike traditional planning models that often assume unlimited resources, our framework directly incorporates budget awareness throughout the tower recommendation pipeline.

We model the deployment constraint using a simple cost equation. If the total budget is denoted by B and the fixed cost of installing a single tower is C, then the upper limit on the number of deployable towers is given by:

Max Sites = $\lfloor B / C \rfloor$

This constraint is not just applied at the final stage of planning but is actively used to guide earlier stages, particularly during the clustering process. When the number of proposed clusters (each representing a low-coverage region) exceeds the allowable number of installations, we employ a cost-benefit ranking mechanism. Clusters are prioritized based on their expected marginal improvement in coverage per unit cost, ensuring that limited resources yield the highest possible network gains.

Furthermore, the framework supports flexible deployment strategies. For example, if placing a tower at a cluster centroid proves insufficient, the system can consider distributing towers across multiple critical points within the cluster. These options are evaluated based on their overall impact relative to budget consumption.

Moreover, this cost-aware mechanism is embedded within the model's iterative feedback loop. As each new tower is virtually "added," the system recalculates coverage levels, identifies remaining gaps, and reprioritizes candidate sites based on both technical benefits and remaining budget. This dynamic adjustment allows for more realistic and adaptive planning, particularly in diverse terrains or markets with varying levels of financial investment.

In summary, by integrating budget constraints directly into the tower recommendation and clustering process, our framework supports strategic decision-making that balances technical performance with financial feasibility. This enables telecom operators to make the most efficient use of available resources while expanding network coverage in a scalable and cost-effective manner.

## VII. Simulation and technical difficulties

Although the proposed framework offers a comprehensive and practical approach to network capacity expansion, simulating the full system proved to be challenging in our current research setup. While the methodology is theoretically sound and modular in design, several real-world constraints and technical limitations restricted our ability to fully implement and validate the model through end-to-end simulations.

One of the main challenges was the integration of heterogeneous data sources. The model requires multiple types of input data, such as drive test logs, performance management counters, subscriber density, terrain information, and existing tower configurations, at a sufficiently granular resolution. In practice, obtaining such data in a unified, consistent, and high-quality format is non-trivial. Many datasets are incomplete, inconsistently formatted, or geographically misaligned, making it difficult to prepare a clean and reliable training and evaluation pipeline.

Another significant challenge was modeling geographic features and their impact on radio propagation. Accurately representing physical elements such as water bodies, high-rise buildings, elevation changes, and forested areas requires detailed geospatial data and advanced simulation tools. Incorporating these features into signal prediction models with high fidelity remains computationally intensive and data-dependent. In our case, we lacked access to fine-grained terrain and urban infrastructure datasets that would allow realistic simulation of signal behavior in complex environments.

Moreover, simulating budget-constrained deployment strategies across large grids introduced additional complexity. The iterative planning loop, where tower placements are updated and coverage is recalculated at each step, requires fast and accurate inference over high-dimensional spatial data. While feasible in principle, this process demands substantial computational resources and robust data pipelines, which were beyond the scope of our current experimental infrastructure.

Finally, the dynamic nature of real-world telecom deployments, such as changes in subscriber behavior, environmental conditions, and network load, adds further complexity that is difficult to capture in a static simulation. While our model is designed to accommodate such variability, reproducing it accurately in a simulated environment would require time-series data and long-term field measurements, which were not available at the time of this study.

Despite these challenges, the theoretical design and modular structure of the framework remain sound. Each component, including the deep learning-based coverage predictor and the clustering-based site recommender, has been individually validated in prior research or internal prototypes. We are confident that with access to richer datasets and appropriate simulation tools, the full framework can be implemented and tested in real-world or semi-synthetic scenarios.

In future work, we aim to overcome these technical barriers by collaborating with telecom operators to access live network data and geospatial maps. We also plan to explore hybrid simulation environments that combine synthetic data with real-world measurements to validate the framework's effectiveness in diverse deployment settings.

## VIII. Discussion

Expanding mobile network coverage is a complex task that involves not only technical challenges but also business and environmental considerations. Traditional network planning methods such as manual drive tests, site surveys, and heuristic rules are still widely used in the industry. However, these approaches are often time-consuming, labor intensive, and limited in their ability to adapt to diverse real-world conditions. They typically do not take into account key factors such as geographic obstacles, changing subscriber distribution, or financial constraints, all of which are essential for effective long-term planning.

Our proposed framework addresses these gaps by introducing a machine learning-based approach that learns from real data to predict signal coverage and recommend suitable locations for new towers. By using a deep neural network trained on features like terrain, subscriber density, and existing infrastructure, the system can make more informed decisions about where coverage is lacking and how to improve it. This predictive model captures complex relationships that simple rules or static thresholds cannot.

In addition, the use of clustering algorithms to identify low-coverage regions allows the system to group nearby problem areas and suggest tower placements at the most effective central locations. By considering cost constraints, such as the maximum number of towers that can be installed within a fixed budget, the model ensures that recommendations are both technically sound and financially practical. This budget-aware design is particularly valuable for operators who must justify capital expenditures with expected returns on investment.

An important advantage of our framework is its iterative design. After each round of planning and simulated Anamandra deployment, the system reevaluates the coverage landscape and updates its recommendations. This step-by-step process reflects how network expansion happens in practice and allows for adaptive decision-making based on the latest data.

Another key strength is the model's robustness to incomplete or imperfect data. In many real-world settings, especially in remote or under-mapped areas, some data may be missing or unreliable. Our method incorporates preprocessing and imputation techniques to handle these gaps and maintain reliable performance.

Overall, this work moves beyond manual and assumption-heavy planning methods by offering a data-driven, scalable, and adaptive solution. By combining signal prediction, spatial analytics, and financial planning into a single pipeline, it provides a more accurate and efficient tool for mobile network expansion. This integrated approach supports telecom operators in making better, faster decisions while reducing reliance on exhaustive field testing and manual analysis.

## IX. Conclusion

This paper proposed an AI-driven framework for mobile network capacity expansion, combining predictive modeling, spatial clustering, and cost-aware optimization. In contrast to traditional manual and heuristic-based methods, our system identifies low-coverage regions and recommends feasible tower placements using a deep neural network trained on geographic, demographic, and infrastructural features. A clustering-based module, guided by budget constraints, selects optimal candidate sites, enabling iterative planning that mirrors real-world network deployment cycles.

The approach advances both the theory and practice of network planning by modeling complex real-world interactions while reducing reliance on drive tests and manual tuning. By embedding financial considerations directly into the optimization process, it enables more transparent, adaptive, and scalable decision-making.

Although full simulation was limited by data heterogeneity and modeling challenges, the modular structure and robustness to missing inputs suggest strong applicability. Future work will involve deployment on live data, refinement of clustering strategies, and dynamic modeling of evolving network demand.